\documentclass[draft]{agujournal2019}
\usepackage{url} 
\usepackage{lineno}
\usepackage[inline]{trackchanges} 
\usepackage{soul}
\usepackage{amsmath,bm}
\usepackage{subcaption}

\draftfalse

\journalname{arXiv}

\begin{document}

%
%


\title{A Method for Calculating Attenuation in Creeping Materials}

%
%




\authors{Ron Maor\affil{1}, Nir Z. Badt\affil{1}, Hugo N. Ulloa\affil{1}, and David L. Goldsby\affil{1}}

\affiliation{1}{Department of Earth and Environmental Science, University of Pennsylvania, PA, USA}





\correspondingauthor{Ron Maor}{ronmaor@sas.upenn.edu}



\begin{keypoints}
\item We provide a method to accurately extract phase lag between complex signals
\item The method is evaluated and tested with data from attenuation experiments
\item We discuss the relevance of this method for studying energy dissipation in geophysical systems and the improved reproducibility in attenuation experiments
\end{keypoints}

%
%

%
%


\begin{abstract}
The phase~lag between an applied forcing and a response to that forcing is a fundamental parameter in geophysical signal processing. For solid deforming materials, the phase~lag between an oscillatory applied stress and the resulting strain response encapsulates information about the dynamical behavior of materials and attenuation. The phase lag is not directly measured and must be extracted through multiple steps by carefully comparing two time-series signals. The extracted value of the phase~lag is highly sensitive to the analysis method, and often there are no comparable values to increase confidence in the calculated results. In this study, we propose a method for extracting the phase~lag between two signals when either one or both include an underlying nonlinear trend, which is very common when measuring attenuation in creeping materials. We demonstrate the robustness of the method by analyzing artificial signals with known phases and quantifying their absolute and relative errors. We apply the method to two experimental datasets and compare our results with those of previous studies. 
\end{abstract}

\section*{Plain Language Summary}
Many natural systems are influenced by regular forcing cycles, and their responses to these cycles often show a delay called the phase~lag. This delay happens because several mechanisms respond to the cyclic forcing, and some dissipate energy. For geological materials, measuring the phase~lag at different forcing frequencies helps to understand the properties of the material and which dynamical processes are involved in the response. This is done by creating an attenuation spectrum. However, since several mechanisms may act at once, it can be challenging to untangle the phase~lag caused by a specific forcing frequency. We propose a new method to accurately measure this phase~lag, even when multiple processes are involved in the response of materials to forced oscillations. We demonstrate the method's applicability to both simulated and real laboratory data and provide an open-source algorithm for reproducing our results and using the method in future studies.

%
%

%


%
%
%
%

\section{Introduction}
The viscoelastic response of materials to external forcing includes an immediate elastic response and a time-dependent viscous response. As such, the viscous response may be a combination of multiple mechanisms operating simultaneously, each over a different timescale. In geological materials, viscoelastic behavior is tied to numerous phenomena, such as post-glacial isostatic adjustment, tidal dissipation and post-seismic deformation \cite{karato1990}. Deformation in geological materials occurs over long time periods (e.g., 10\textsuperscript{15}~s for mantle convection), at high homologous temperatures, meaning that these materials exhibit viscous creep in conjunction with viscoelastic behavior \cite{karato2008}. Measuring the viscoelastic response of geological materials in the laboratory is typically done by quantifying the attenuation or internal friction $Q^{-1}$ in forced oscillation experiments performed at elevated pressure and temperature conditions \cite{jackson1993,tan2001,cooper2002,jackson2002,cao2021}. One of the challenges is that tested samples exhibit both long-term creep and viscoelastic behavior, making data interpretation difficult and susceptible to errors in the estimation of $Q^{-1}$.
\par
Additional complications in estimating $Q^{-1}$ from experiments stem from data analysis techniques that are used to fit experimental results. Attenuation is calculated as the tangent of the phase shift ($\phi$) between two signals in forced oscillation experiments \cite{cooper2002, mccarthy2016tidal, cao2021}, typically to the force (or stress) and displacement (strain) signals. Fitting harmonic functions to the stress and strain data entail biases that arise from, for example, assuming that both harmonic signals have the same oscillation frequency, or by detrending a strain signal that also includes a long-term creep component. Even if the spectral separation between the creep behavior and the oscillation frequency is large there is still a problem in fitting multiple parameters for both signals, such as amplitude and phase. Furthermore, because the phase shift of a signal is highly sensitive metric even the length of laboratory-derived data series being fitted to a harmonic function may result in different values of $\phi$.
\par
A common goal of performing attenuation experiments on geological materials is to obtain an attenuation spectrum, which shows the relative energy dissipation in the material at different forcing frequencies. Therefore, there is a need for a robust method to calculate the phase shift at each frequency in a consistent manner, eliminating the additional error inherent in the coefficients of each fit.
\par
We present and test a new method, coined `High-Pass Phase Detection' (HPPD), for calculating the phase shift between two signals (stress and strain) from forced oscillation experiments of creeping materials. This new method employs high-pass signal filtering coupled with the Fast Fourier Transform (FFT) to compute the phase shift through a consistent algorithm, without the need to perform curve-fitting for each signal individually. We show that this new method can consistently recover the phase shift of synthetic signals containing both linear and nonlinear trends. We then apply the new method to experimental data and demonstrate that $Q^{-1}$ can be calculated efficiently and that extracted values are consistent with those of previous studies.

\section{Method: High-Pass Phase Detection (HPPD)}\label{sec:method}

Let us consider a general signal composed of a  sinusoidal part and an underlying creep trend, 
\begin{equation}
    x[t] = c(t) + A \sin(2 \pi f t + \phi) = \bar{c}[t] + \bar{s}[t],
\end{equation}
and its Fourier Transform (FT):
\begin{equation}
    X[\omega] = C[\omega] + S[\omega].
\end{equation}
The phase of the frequency $\omega$ in the FT spectrum is simultaneously influenced by the sinusoidal component and the creep component. However, if there is enough spectral separation between the primary frequencies constituting the sinusoidal component ($S$) and those constituting the creep component ($C$), filtering methods can be applied to remove the effect of creep on the phase associated with the frequency $\omega$.

To achieve a robust separation between the periodic component that carries the information to estimate $Q^{-1}$ and the creep signal, we use a high-pass response filter with a Butterworth design \cite{Oppenheim1982-yv,butterworth2020design}. This approach requires specifying a cutoff frequency for the high-pass filter close to the known frequency of the oscillations. Here, we utilize a second-order Butterworth filter. Note that when applying a digital filter, a spurious transient effect occurs in the initial oscillations of the filtered signal \cite{transientDSP1998introductory}. To address this issue, it is necessary to remove the transient effect from the filtered signal before calculating the phase shift $\phi$.

\subsection{Testing with Synthetic Signals}

In order to test and quantify the error of the filtering method, we construct synthetic signals with both linear and nonlinear trends and rediscover the phase. The generated signals have the following form:
\begin{equation} \label{eq:synsigs}
\begin{aligned}
h_i(t) &=  h_{0}\sin(2\pi f t + \phi_j) + a_i t , \\ 
g_i(t) &=  g_{0}\sin(2\pi f t + \phi_j) + b_i t^2  + a_i t,
\end{aligned}
\end{equation}
where $h_i(t)$ represents a periodic signal with a linear trend, and $g_i(t)$ represents a periodic signal with a nonlinear trend. Both signals have a frequency $f$, a phase $\phi_j$, and coefficients $a_i$ and $b_i$ weighing the trends. We remark that we only deal with the phase and frequency of the signals, hence the units of the other parameters will remain general.

\begin{figure}
\noindent\includegraphics[width=\textwidth]{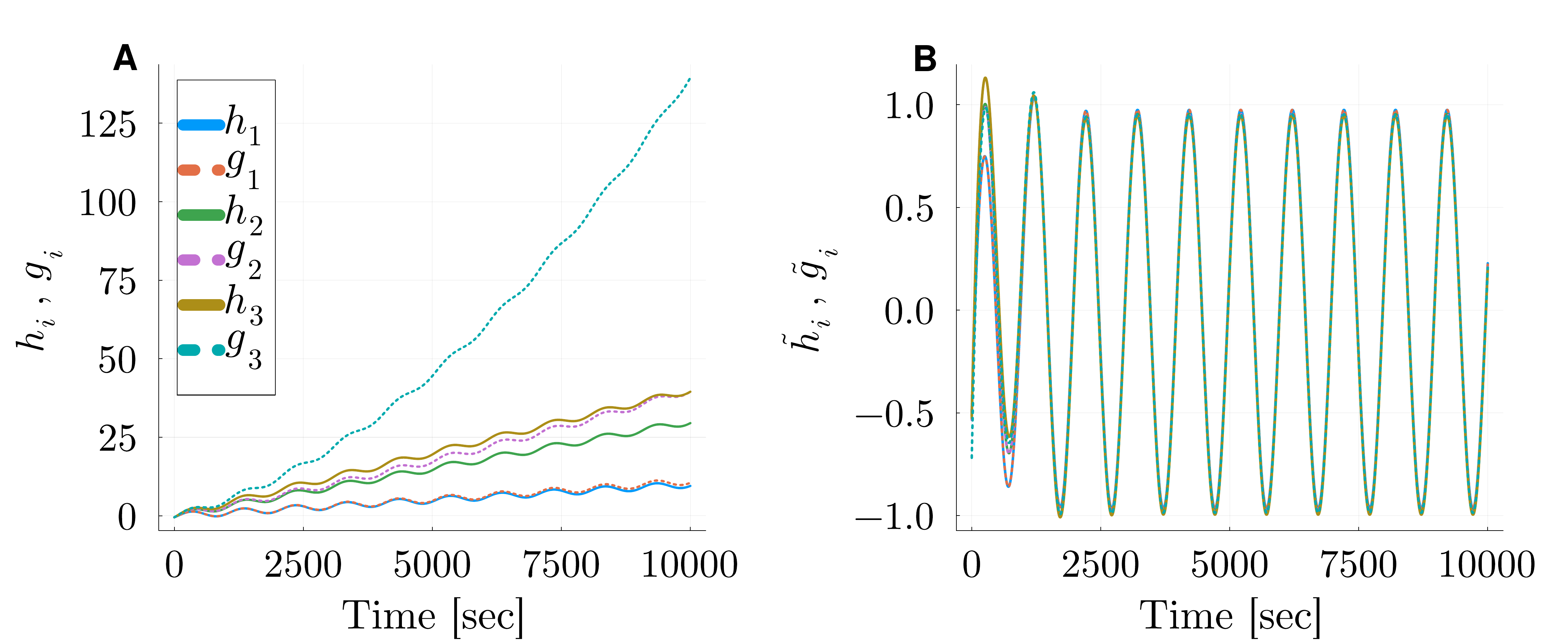}
\caption{(A) The synthetic signals from Equation~\eqref{eq:synsigs}. The sampling frequency is 10~Hz, oscillation frequency is 10\textsuperscript{-3}~Hz, and all the signals have a phase of $\phi= \pi /6$. Dotted lines represent the signals with the nonlinear trend, and solid lines represent the linear trend signals. (B) The same signals after applying the filter. Notice the transient effect at the beginning.}
\label{fig:syn_signals}
\end{figure}

We test the method using phase angles corresponding to the viscoelastic regime ($0< \phi_j < \pi/2$), and choose forcing and sampling frequencies that are common in attenuation experiments on geological materials. Figure \ref{fig:syn_signals}A shows the generated signals that carry the same phase angle. To challenge the method, we deliberately selected a diverse set of trends, spanning a wide range of values that differ by two orders of magnitude.

\begin{table}
\caption{Parameters of the signals in Figure \ref{fig:syn_signals}A and the errors of detecting their phase in Figure \ref{fig:sigserr-rel}A. All signals have a frequency of 10\textsuperscript{-3}~Hz, phase~$\phi$
~=~$\pi/6$, and amplitude $h_0=g_0=1$. The calibration bias is $\sim$0.756}
\centering
\begin{tabular}{l c c c}
\hline
Signal & $a_{i}$ & $b_{i}$ & Absolute error [\%] \\
\hline
$h_1$ & $10^{-3}$ & - & $1.0 \times 10^{-7}$ \\
$h_2$ & $3 \times 10^{-3}$ & - & $8.5 \times 10^{-5}$ \\
$h_3$ & $4 \times 10^{-3}$ & - & $1.27\times 10^{-4}$ \\
$g_1$ & $10^{-3}$ & $10^{-8}$ & $8.19 \times 10^{-5}$ \\
$g_2$ & $3 \times 10^{-3}$ & $10^{-7}$ & $7.67 \times 10^{-3}$ \\
$g_3$ & $4 \times 10^{-3}$ & $10^{-6}$ & $4.54 \times 10^{-3}$ \\
\hline
\label{tab:signals}
\end{tabular}
\end{table}

The underlying trends from all the signals illustrated in Figure~\ref{fig:syn_signals}A are filtered out using a high-pass filter with a second-order Butterworth design. We set the cutoff frequency of the high-pass filter to be half of the oscillation frequency. The results of the filtering are presented in Figure~\ref{fig:syn_signals}B. As shown in the figure, the transient effect of the filter affects the first three oscillations of the signals, which need to be removed from the signals before calculating the phase $\phi$. Because of this spurious transient behavior resulting from the filtering process, having a sufficient number of oscillations in the data is critical when using this method.

\begin{figure}
\noindent\includegraphics[width=\textwidth]{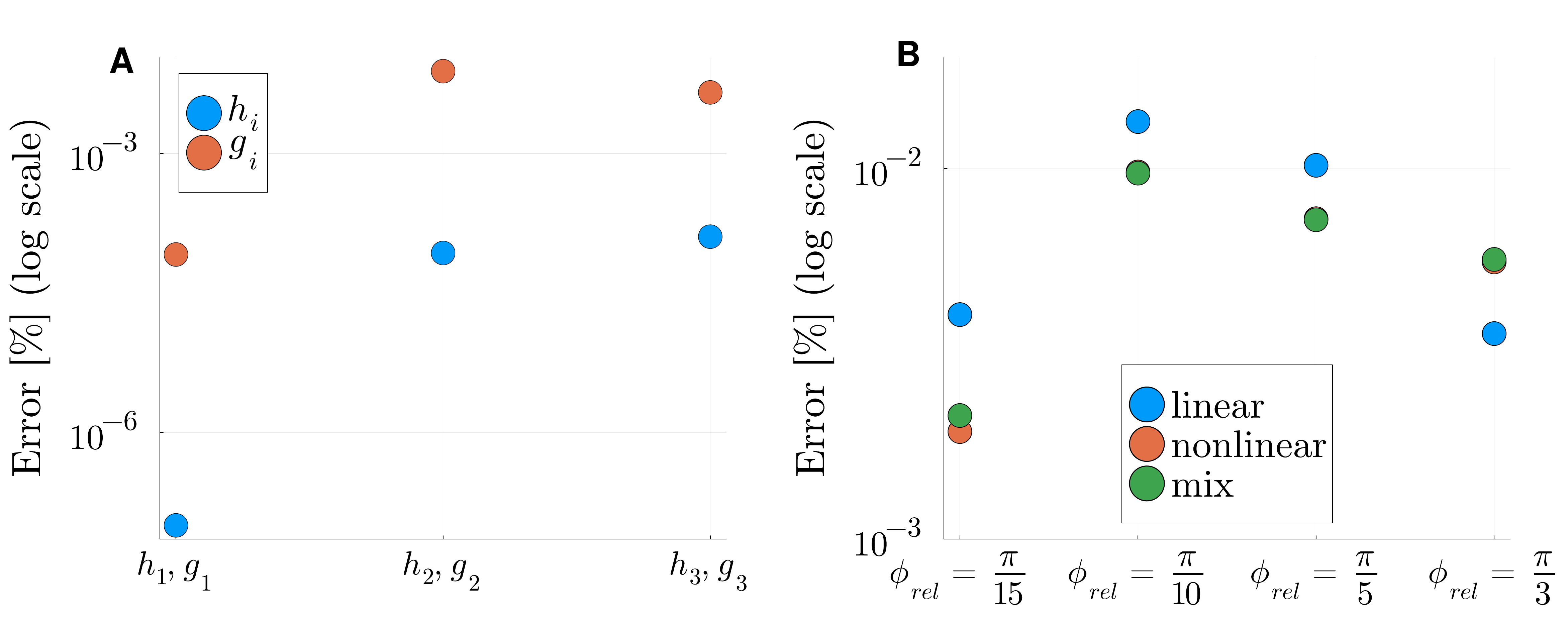}
\caption{(A) The errors between the extracted phase lags for each signal using the filtering method to the known value from constructing the signals. All the errors are below 0.02\% error
 (B) Relative phase lag errors between linear trends, nonlinear trends and a mixture of one linear trend and one nonlinear trend. All the errors are below 0.1\% error.}
\label{fig:sigserr-rel}
\end{figure}

After removing the initial transient oscillations from the filtered signals, we use the Fast Fourier Transform (FFT) to calculate the phase of the dominant frequency in each signal. The generated signals allow us to derive the absolute value of the phase, which is known \textit{a~priori} from constructing the signals in  Equation~\eqref{eq:synsigs} with paramaters summarized in Table~\ref{tab:signals}. However, in order to obtain the exact absolute value of the phase, we need to remove the bias introduced by the filter. This bias is a single value attributed to the design of the filter and is consistent across all processed signals. Following this calibration, we calculate the error for each signal, as presented in Figure \ref{fig:sigserr-rel}A and summarized in Table \ref{tab:signals}.

When analyzing experimental datasets, the desired parameter is the relative phase angle between the stress and strain signals in each experiment, hence there is no need to calibrate the bias of the filter. To mimic this scenario, we repeat the process described above but now calculate the relative phase angles between different sets of signals: two with linear trends, two with nonlinear trends, and a mixture of a linear trend signal and a nonlinear trend signal. The errors of the relative phase lags are reported in Figure \ref{fig:sigserr-rel}B — all of them are below 0.1\%. 

\section{Application: Attenuation Experiments on Halite and PMMA}

\begin{figure}
\noindent\includegraphics[width=\textwidth]{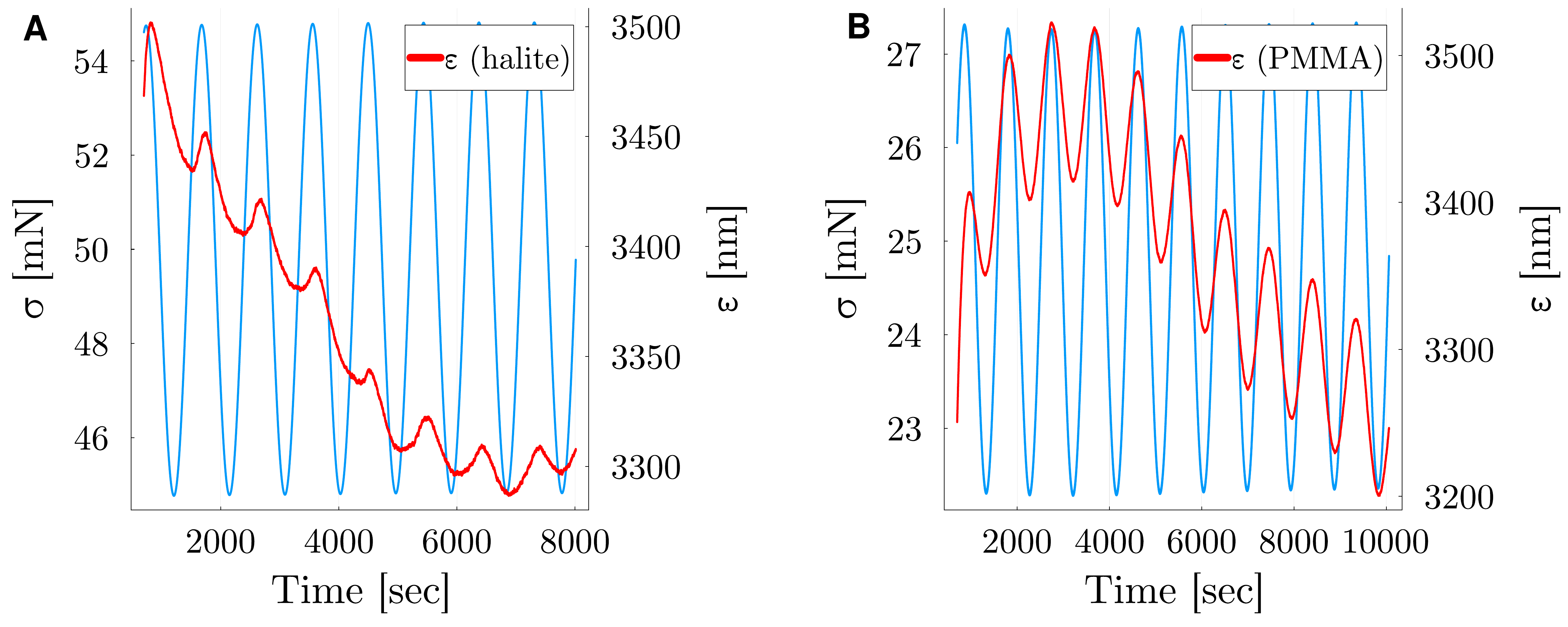}
\caption{Raw data from the nanoindentation experiments. The forcing frequency is set to 10\textsuperscript{-3} Hz, and the sampling frequency is 1 Hz. $\sigma$ represents the stress with units of milli-Newtons, and $\varepsilon$ represents the strain with units of nanometers. (A) Halite sample. (B) PMMA sample.
}
\label{fig:raw}
\end{figure}

We test the method using data from attenuation experiments performed with nanoindentation \cite{badt_2024_GRL}. This dataset is part of a recent study by \cite{badt_2024_GRL} who introduced an experimental technique for measuring the attenuation of halite, olivine, quartz, PMMA and indium by conducting forced oscillation experiments \cite{cooper2002, jackson1993, cao2021} with a nanoindenter. Figure \ref{fig:raw} includes samples of raw data from oscillations at 10\textsuperscript{-3} Hz where nonlinear creep trends are present in the strain signals. \citeA{badt_2024_GRL} reported very good agreement between their results and attenuation spectra obtained in previous experiments using different experimental apparatuses. However, in their analysis, \citeA{badt_2024_GRL} estimate the phase shift using a least-squares algorithm, a common approach in previous attenuation studies \cite{cao2021, mccarthy2016tidal, tan2001}.

\begin{figure}
\includegraphics[width=\textwidth]{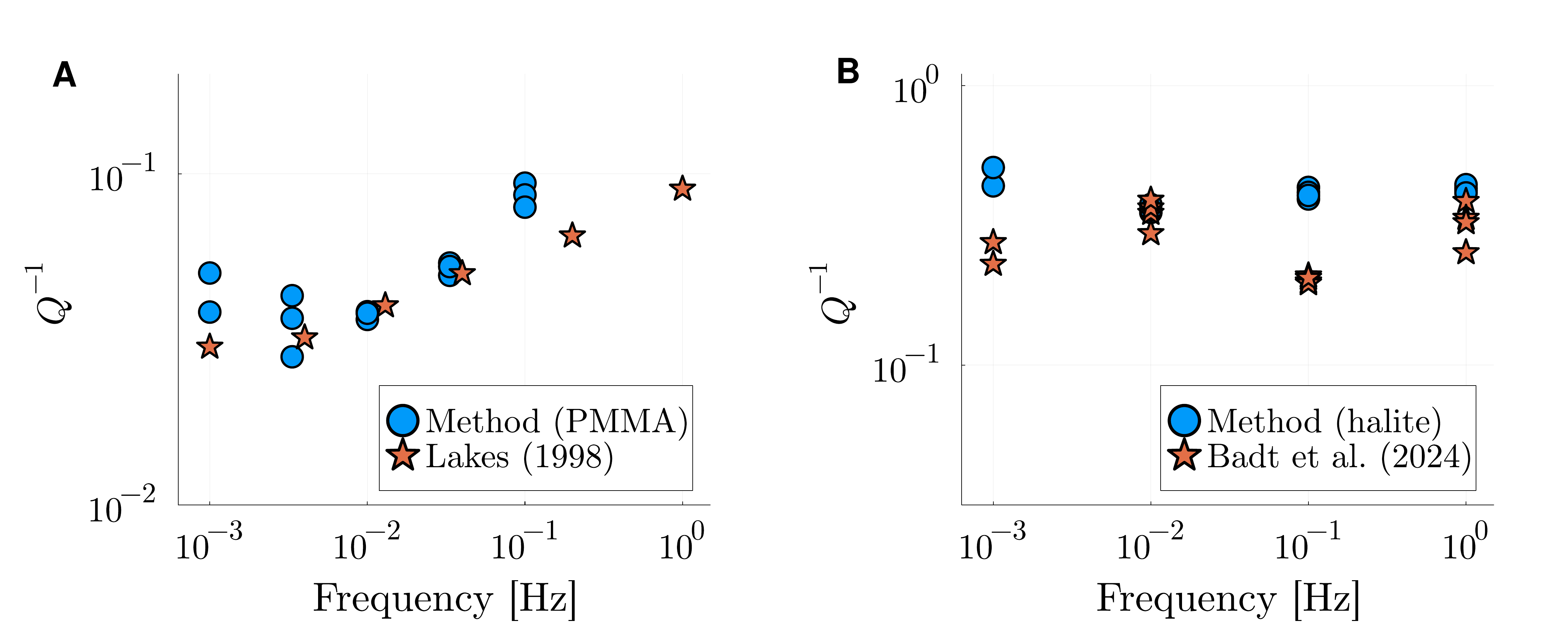}
\caption{Log-scale attenuation spectra vs forcing frequency. (A) Attenuation experiments on (A) PMMA and (B) halite using nanoindentation. Blue circles show attenuation results vs frequency calculated using the method introduced in Section~\ref{sec:method}. Red stars show attenuation results from Lakes~(1998) and Badt~et~al.~(2024), respectively.}
\label{fig:Q}
\end{figure}

Figure~\ref{fig:Q}A shows results from PMMA samples compared to previously published data for PMMA in \cite{lakes1998}. Each blue marker represents a single experiment processed using the method proposed in Section~\ref{sec:method}. All samples in the figure exhibited a creep trend in the strain signal, while the oscillatory stress signal was smooth and did not require filtering. Each strain signal was filtered using the proposed method, and the transient effect of the filter was removed before calculating the phase between the stress and strain signals. The duration of the transient effect varied among the samples; however, none exceeded four consecutive oscillations. Figure~\ref{fig:Q}B displays the attenuation spectrum calculated for halite. The tests on the halite samples showed a pronounced nonlinear creep trend in all strain signals, but the transient effect did not exceed the length of four oscillations across the spectrum. The deviation of $Q^{-1}$ results in PMMA obtained by nanoindentation from the reported values from \cite{lakes1998} at a frequency $0.1$~Hz (Figure~\ref{fig:Q}A) is due to inherent damping in the nanoindenter \cite{badt_2024_GRL}.

\section{Discussion}

Previous studies on attenuation have used nonlinear square fit methods to extract the phase lag \cite{mccarthy2016tidal, QNLStakei2014temperature}, often employing the Levenberg–Marquardt algorithm \cite{levenberg1944method, pujol2007solution}. This approach requires several steps that are absent in the method presented here. First, there is a need to decide on the functional form of the trend in the signal, such as a linear trend, second-order polynomial, etc. This is usually done by trial and error. Second, the algorithm must be provided with an initial guess for each parameter. This is not straightforward in cases of strong nonlinear creep trends, often occurring at low frequencies, as shown for halite in Figure~\ref{fig:Q}B. Numerical algorithms like Levenberg–Marquardt are very sensitive to initial conditions, adding more complexity and uncertainties to compute the phase lag. In the case of nonlinear trends, there might be six or more parameters to fit, making it difficult to estimate how the error of one parameter is compensated by the others. Moreover, the attenuation is defined as $Q^{-1} = \tan{(\phi)}$, which makes it very sensitive to errors in $\phi$ when values reach $\pi/6$ (i.e., where the linear approximation of $\tan(x)\approx x$ is not valid).

In the method demonstrated here, the signals are processed in a non-iterative way, focusing exclusively on the required phase parameter. This simplifies and strengthen significantly the data analysis process for attenuation experiments and increases the confidence of the calculated values. The only requirement is that the trend of the signal has enough spectral separation from the forced oscillations. This is generally the case since the trend is considered as a very low frequency background signal that does not show a periodic behavior in the entire length of the time series. This fact enables us to detect the phase corresponding to the oscillations of the experiment using FFT accurately. In addition, the process of detecting the phase lag in this study can be easily automated, which is useful when analyzing large datasets or when using numerical models. Rheological models for the upper mantle often involve a power law formulation that leads to nonlinear viscoelasticity \cite{jain2019global,freed2004evidence,masuti2023transient} and by applying the suggested method, it is possible to use outputs from time-domain power law models to construct detailed attenuation spectra.

Besides attenuation experiments, the phase lag between a periodic forcing and the material response is ubiquitous in fluid systems, for instance, in the circadian radiative forcing of solar radiation in ice-covered waters like lakes. This under-ice radiative heating leads to a periodic yet transient diurnal convective regime over time due to fluid mixing and the nonlinear equation of state of water \cite{ulloa2019differential}. The result is a periodic signal in the under-ice water temperature and velocity field with a nonlinear trend, whose actual phase lag response provides information on the transformation rate of available potential energy and kinetic energy in the system \cite{winters2019energetics}. The proposed method can be readily applied to determine such a phase lag. Other examples include the phase lag between temperature and density in the thermosphere \cite{kodikara2019density} and the oceanic response to wind forcing \cite{gille2001antarctic,zhang2021spurious}.

\subsection{Conclusions}

Currently, there is no consensus on a universally accepted method to determine the `phase~lag' $\phi$ between stress and strain signals in attenuation experiments on geological materials (e.g., Cao~et.~al,~2021). Moreover, the limited amount of data existing at low frequencies ($\leq$ 1 Hz) makes it difficult to assess if the measured $\phi$ and $Q^{-1}$ are consistent. The proposed method (HPPD) for calculating the phase~lag presented in this study proves to be effective and relevant for quantifying attenuation in geological materials. We hope the method presented will be useful for future experiments, rendering phase~lag extraction more automated and reproducible.

\section*{Open Research Section}
The data to produce the figures in the article, along with the algorithm to calculate the phase lags are available as Julia code \cite{bezanson2017julia} on github.


%
%


%
%
%
%
%

\bibliography{ref}

\begin{thebibliography}{}

\bibitem [\protect \citeauthoryear {%
Badt%
, Maor%
\BCBL {}\ \BBA {} Goldsby%
}{%
Badt%
\ \protect \BOthers {.}}{%
{\protect \APACyear {2024}}%
}]{%
badt_2024_GRL}
\APACinsertmetastar {%
badt_2024_GRL}%
\begin{APACrefauthors}%
Badt, N.%
, Maor, R.%
\BCBL {}\ \BBA {} Goldsby, D.%
\end{APACrefauthors}%
\unskip\
\newblock
\APACrefYearMonthDay{2024}{}{}.
\newblock
{\BBOQ}\APACrefatitle {A nanoindentation study of attenuation in geological materials} {A nanoindentation study of attenuation in geological materials}.{\BBCQ}
\newblock
\APACjournalVolNumPages{Under review in Geophys. Res. Lett.}{}{}{}.
\PrintBackRefs{\CurrentBib}

\bibitem [\protect \citeauthoryear {%
Bezanson%
, Edelman%
, Karpinski%
\BCBL {}\ \BBA {} Shah%
}{%
Bezanson%
\ \protect \BOthers {.}}{%
{\protect \APACyear {2017}}%
}]{%
bezanson2017julia}
\APACinsertmetastar {%
bezanson2017julia}%
\begin{APACrefauthors}%
Bezanson, J.%
, Edelman, A.%
, Karpinski, S.%
\BCBL {}\ \BBA {} Shah, V\BPBI B.%
\end{APACrefauthors}%
\unskip\
\newblock
\APACrefYearMonthDay{2017}{}{}.
\newblock
{\BBOQ}\APACrefatitle {Julia: A fresh approach to numerical computing} {Julia: A fresh approach to numerical computing}.{\BBCQ}
\newblock
\APACjournalVolNumPages{SIAM review}{59}{1}{65--98}.
\newblock
\begin{APACrefURL} \url{https://doi.org/10.1137/141000671} \end{APACrefURL}
\PrintBackRefs{\CurrentBib}

\bibitem [\protect \citeauthoryear {%
Cao%
, Hansen%
, Thom%
\BCBL {}\ \BBA {} Wallis%
}{%
Cao%
\ \protect \BOthers {.}}{%
{\protect \APACyear {2021}}%
}]{%
cao2021}
\APACinsertmetastar {%
cao2021}%
\begin{APACrefauthors}%
Cao, R.%
, Hansen, L\BPBI N.%
, Thom, C\BPBI A.%
\BCBL {}\ \BBA {} Wallis, D.%
\end{APACrefauthors}%
\unskip\
\newblock
\APACrefYearMonthDay{2021}{}{}.
\newblock
{\BBOQ}\APACrefatitle {An apparatus for measuring nonlinear viscoelasticity of minerals at high temperature} {An apparatus for measuring nonlinear viscoelasticity of minerals at high temperature}.{\BBCQ}
\newblock
\APACjournalVolNumPages{Review of Scientific Instruments}{92}{7}{}.
\PrintBackRefs{\CurrentBib}

\bibitem [\protect \citeauthoryear {%
Cooper%
}{%
Cooper%
}{%
{\protect \APACyear {2002}}%
}]{%
cooper2002}
\APACinsertmetastar {%
cooper2002}%
\begin{APACrefauthors}%
Cooper, R\BPBI F.%
\end{APACrefauthors}%
\unskip\
\newblock
\APACrefYearMonthDay{2002}{}{}.
\newblock
{\BBOQ}\APACrefatitle {Seismic wave attenuation: Energy dissipation in viscoelastic crystalline solids} {Seismic wave attenuation: Energy dissipation in viscoelastic crystalline solids}.{\BBCQ}
\newblock
\APACjournalVolNumPages{Reviews in mineralogy and geochemistry}{51}{1}{253--290}.
\PrintBackRefs{\CurrentBib}

\bibitem [\protect \citeauthoryear {%
Freed%
\ \BBA {} B{\"u}rgmann%
}{%
Freed%
\ \BBA {} B{\"u}rgmann%
}{%
{\protect \APACyear {2004}}%
}]{%
freed2004evidence}
\APACinsertmetastar {%
freed2004evidence}%
\begin{APACrefauthors}%
Freed, A\BPBI M.%
\BCBT {}\ \BBA {} B{\"u}rgmann, R.%
\end{APACrefauthors}%
\unskip\
\newblock
\APACrefYearMonthDay{2004}{}{}.
\newblock
{\BBOQ}\APACrefatitle {Evidence of power-law flow in the Mojave desert mantle} {Evidence of power-law flow in the mojave desert mantle}.{\BBCQ}
\newblock
\APACjournalVolNumPages{Nature}{430}{6999}{548--551}.
\PrintBackRefs{\CurrentBib}

\bibitem [\protect \citeauthoryear {%
Gille%
, Stevens%
, Tokmakian%
\BCBL {}\ \BBA {} Heywood%
}{%
Gille%
\ \protect \BOthers {.}}{%
{\protect \APACyear {2001}}%
}]{%
gille2001antarctic}
\APACinsertmetastar {%
gille2001antarctic}%
\begin{APACrefauthors}%
Gille, S\BPBI T.%
, Stevens, D\BPBI P.%
, Tokmakian, R\BPBI T.%
\BCBL {}\ \BBA {} Heywood, K\BPBI J.%
\end{APACrefauthors}%
\unskip\
\newblock
\APACrefYearMonthDay{2001}{}{}.
\newblock
{\BBOQ}\APACrefatitle {Antarctic Circumpolar Current response to zonally averaged winds} {Antarctic circumpolar current response to zonally averaged winds}.{\BBCQ}
\newblock
\APACjournalVolNumPages{Journal of Geophysical Research: Oceans}{106}{C2}{2743--2759}.
\PrintBackRefs{\CurrentBib}

\bibitem [\protect \citeauthoryear {%
Jackson%
, Fitz~Gerald%
, Faul%
\BCBL {}\ \BBA {} Tan%
}{%
Jackson%
\ \protect \BOthers {.}}{%
{\protect \APACyear {2002}}%
}]{%
jackson2002}
\APACinsertmetastar {%
jackson2002}%
\begin{APACrefauthors}%
Jackson, I.%
, Fitz~Gerald, J\BPBI D.%
, Faul, U\BPBI H.%
\BCBL {}\ \BBA {} Tan, B\BPBI H.%
\end{APACrefauthors}%
\unskip\
\newblock
\APACrefYearMonthDay{2002}{}{}.
\newblock
{\BBOQ}\APACrefatitle {Grain-size-sensitive seismic wave attenuation in polycrystalline olivine} {Grain-size-sensitive seismic wave attenuation in polycrystalline olivine}.{\BBCQ}
\newblock
\APACjournalVolNumPages{Journal of Geophysical Research: Solid Earth}{107}{B12}{ECV--5}.
\PrintBackRefs{\CurrentBib}

\bibitem [\protect \citeauthoryear {%
Jackson%
\ \BBA {} Paterson%
}{%
Jackson%
\ \BBA {} Paterson%
}{%
{\protect \APACyear {1993}}%
}]{%
jackson1993}
\APACinsertmetastar {%
jackson1993}%
\begin{APACrefauthors}%
Jackson, I.%
\BCBT {}\ \BBA {} Paterson, M.%
\end{APACrefauthors}%
\unskip\
\newblock
\APACrefYearMonthDay{1993}{}{}.
\newblock
{\BBOQ}\APACrefatitle {A high-pressure, high-temperature apparatus for studies of seismic wave dispersion and attenuation} {A high-pressure, high-temperature apparatus for studies of seismic wave dispersion and attenuation}.{\BBCQ}
\newblock
\APACjournalVolNumPages{Pure and Applied Geophysics}{141}{}{445--466}.
\PrintBackRefs{\CurrentBib}

\bibitem [\protect \citeauthoryear {%
Jain%
, Korenaga%
\BCBL {}\ \BBA {} Karato%
}{%
Jain%
\ \protect \BOthers {.}}{%
{\protect \APACyear {2019}}%
}]{%
jain2019global}
\APACinsertmetastar {%
jain2019global}%
\begin{APACrefauthors}%
Jain, C.%
, Korenaga, J.%
\BCBL {}\ \BBA {} Karato, S\BHBI i.%
\end{APACrefauthors}%
\unskip\
\newblock
\APACrefYearMonthDay{2019}{}{}.
\newblock
{\BBOQ}\APACrefatitle {Global analysis of experimental data on the rheology of olivine aggregates} {Global analysis of experimental data on the rheology of olivine aggregates}.{\BBCQ}
\newblock
\APACjournalVolNumPages{Journal of Geophysical Research: Solid Earth}{124}{1}{310--334}.
\PrintBackRefs{\CurrentBib}

\bibitem [\protect \citeauthoryear {%
Karato%
}{%
Karato%
}{%
{\protect \APACyear {2008}}%
}]{%
karato2008}
\APACinsertmetastar {%
karato2008}%
\begin{APACrefauthors}%
Karato, S\BHBI i.%
\end{APACrefauthors}%
\unskip\
\newblock
\APACrefYearMonthDay{2008}{}{}.
\newblock
{\BBOQ}\APACrefatitle {Deformation of earth materials} {Deformation of earth materials}.{\BBCQ}
\newblock
\APACjournalVolNumPages{An introduction to the rheology of Solid Earth}{463}{}{}.
\PrintBackRefs{\CurrentBib}

\bibitem [\protect \citeauthoryear {%
Karato%
\ \BBA {} Spetzler%
}{%
Karato%
\ \BBA {} Spetzler%
}{%
{\protect \APACyear {1990}}%
}]{%
karato1990}
\APACinsertmetastar {%
karato1990}%
\begin{APACrefauthors}%
Karato, S\BHBI i.%
\BCBT {}\ \BBA {} Spetzler, H.%
\end{APACrefauthors}%
\unskip\
\newblock
\APACrefYearMonthDay{1990}{}{}.
\newblock
{\BBOQ}\APACrefatitle {Defect microdynamics in minerals and solid-state mechanisms of seismic wave attenuation and velocity dispersion in the mantle} {Defect microdynamics in minerals and solid-state mechanisms of seismic wave attenuation and velocity dispersion in the mantle}.{\BBCQ}
\newblock
\APACjournalVolNumPages{Reviews of Geophysics}{28}{4}{399--421}.
\PrintBackRefs{\CurrentBib}

\bibitem [\protect \citeauthoryear {%
Kodikara%
, Carter%
, Norman%
\BCBL {}\ \BBA {} Zhang%
}{%
Kodikara%
\ \protect \BOthers {.}}{%
{\protect \APACyear {2019}}%
}]{%
kodikara2019density}
\APACinsertmetastar {%
kodikara2019density}%
\begin{APACrefauthors}%
Kodikara, T.%
, Carter, B.%
, Norman, R.%
\BCBL {}\ \BBA {} Zhang, K.%
\end{APACrefauthors}%
\unskip\
\newblock
\APACrefYearMonthDay{2019}{}{}.
\newblock
{\BBOQ}\APACrefatitle {Density-temperature synchrony in the hydrostatic thermosphere} {Density-temperature synchrony in the hydrostatic thermosphere}.{\BBCQ}
\newblock
\APACjournalVolNumPages{Journal of Geophysical Research: Space Physics}{124}{1}{674--699}.
\PrintBackRefs{\CurrentBib}

\bibitem [\protect \citeauthoryear {%
Lakes%
}{%
Lakes%
}{%
{\protect \APACyear {1998}}%
}]{%
lakes1998}
\APACinsertmetastar {%
lakes1998}%
\begin{APACrefauthors}%
Lakes, R\BPBI S.%
\end{APACrefauthors}%
\unskip\
\newblock
\APACrefYear{1998}.
\newblock
\APACrefbtitle {Viscoelastic Solids (1998)} {Viscoelastic solids (1998)}.
\newblock
\APACaddressPublisher{}{CRC press}.
\PrintBackRefs{\CurrentBib}

\bibitem [\protect \citeauthoryear {%
Levenberg%
}{%
Levenberg%
}{%
{\protect \APACyear {1944}}%
}]{%
levenberg1944method}
\APACinsertmetastar {%
levenberg1944method}%
\begin{APACrefauthors}%
Levenberg, K.%
\end{APACrefauthors}%
\unskip\
\newblock
\APACrefYearMonthDay{1944}{}{}.
\newblock
{\BBOQ}\APACrefatitle {A method for the solution of certain non-linear problems in least squares} {A method for the solution of certain non-linear problems in least squares}.{\BBCQ}
\newblock
\APACjournalVolNumPages{Quarterly of applied mathematics}{2}{2}{164--168}.
\PrintBackRefs{\CurrentBib}

\bibitem [\protect \citeauthoryear {%
Lynn%
\ \BBA {} Fuerst%
}{%
Lynn%
\ \BBA {} Fuerst%
}{%
{\protect \APACyear {1998}}%
}]{%
transientDSP1998introductory}
\APACinsertmetastar {%
transientDSP1998introductory}%
\begin{APACrefauthors}%
Lynn, P\BPBI A.%
\BCBT {}\ \BBA {} Fuerst, W.%
\end{APACrefauthors}%
\unskip\
\newblock
\APACrefYear{1998}.
\newblock
\APACrefbtitle {Introductory digital signal processing with computer applications} {Introductory digital signal processing with computer applications}.
\newblock
\APACaddressPublisher{}{John Wiley \& Sons}.
\PrintBackRefs{\CurrentBib}

\bibitem [\protect \citeauthoryear {%
Masuti%
, Muto%
\BCBL {}\ \BBA {} Rybacki%
}{%
Masuti%
\ \protect \BOthers {.}}{%
{\protect \APACyear {2023}}%
}]{%
masuti2023transient}
\APACinsertmetastar {%
masuti2023transient}%
\begin{APACrefauthors}%
Masuti, S.%
, Muto, J.%
\BCBL {}\ \BBA {} Rybacki, E.%
\end{APACrefauthors}%
\unskip\
\newblock
\APACrefYearMonthDay{2023}{}{}.
\newblock
{\BBOQ}\APACrefatitle {Transient creep of quartz and granulite at high temperature under wet conditions} {Transient creep of quartz and granulite at high temperature under wet conditions}.{\BBCQ}
\newblock
\APACjournalVolNumPages{Journal of Geophysical Research: Solid Earth}{128}{10}{e2023JB027762}.
\PrintBackRefs{\CurrentBib}

\bibitem [\protect \citeauthoryear {%
McCarthy%
\ \BBA {} Cooper%
}{%
McCarthy%
\ \BBA {} Cooper%
}{%
{\protect \APACyear {2016}}%
}]{%
mccarthy2016tidal}
\APACinsertmetastar {%
mccarthy2016tidal}%
\begin{APACrefauthors}%
McCarthy, C.%
\BCBT {}\ \BBA {} Cooper, R\BPBI F.%
\end{APACrefauthors}%
\unskip\
\newblock
\APACrefYearMonthDay{2016}{}{}.
\newblock
{\BBOQ}\APACrefatitle {Tidal dissipation in creeping ice and the thermal evolution of Europa} {Tidal dissipation in creeping ice and the thermal evolution of europa}.{\BBCQ}
\newblock
\APACjournalVolNumPages{Earth and Planetary Science Letters}{443}{}{185--194}.
\PrintBackRefs{\CurrentBib}

\bibitem [\protect \citeauthoryear {%
Oppenheim%
\ \BBA {} Willsky%
}{%
Oppenheim%
\ \BBA {} Willsky%
}{%
{\protect \APACyear {1982}}%
}]{%
Oppenheim1982-yv}
\APACinsertmetastar {%
Oppenheim1982-yv}%
\begin{APACrefauthors}%
Oppenheim, A\BPBI V.%
\BCBT {}\ \BBA {} Willsky, A\BPBI S.%
\end{APACrefauthors}%
\unskip\
\newblock
\APACrefYear{1982}.
\newblock
\APACrefbtitle {Signals and Systems} {Signals and systems}.
\newblock
\APACaddressPublisher{Old Tappan, NJ}{Prentice Hall}.
\PrintBackRefs{\CurrentBib}

\bibitem [\protect \citeauthoryear {%
Pujol%
}{%
Pujol%
}{%
{\protect \APACyear {2007}}%
}]{%
pujol2007solution}
\APACinsertmetastar {%
pujol2007solution}%
\begin{APACrefauthors}%
Pujol, J.%
\end{APACrefauthors}%
\unskip\
\newblock
\APACrefYearMonthDay{2007}{}{}.
\newblock
{\BBOQ}\APACrefatitle {The solution of nonlinear inverse problems and the Levenberg-Marquardt method} {The solution of nonlinear inverse problems and the levenberg-marquardt method}.{\BBCQ}
\newblock
\APACjournalVolNumPages{Geophysics}{72}{4}{W1--W16}.
\PrintBackRefs{\CurrentBib}

\bibitem [\protect \citeauthoryear {%
Shouran%
\ \BBA {} Elgamli%
}{%
Shouran%
\ \BBA {} Elgamli%
}{%
{\protect \APACyear {2020}}%
}]{%
butterworth2020design}
\APACinsertmetastar {%
butterworth2020design}%
\begin{APACrefauthors}%
Shouran, M.%
\BCBT {}\ \BBA {} Elgamli, E.%
\end{APACrefauthors}%
\unskip\
\newblock
\APACrefYearMonthDay{2020}{}{}.
\newblock
{\BBOQ}\APACrefatitle {Design and implementation of Butterworth filter} {Design and implementation of butterworth filter}.{\BBCQ}
\newblock
\APACjournalVolNumPages{International Journal of Innovative Research in Science Engineering and Technology}{9}{9}{7975--7983}.
\PrintBackRefs{\CurrentBib}

\bibitem [\protect \citeauthoryear {%
Takei%
, Karasawa%
\BCBL {}\ \BBA {} Yamauchi%
}{%
Takei%
\ \protect \BOthers {.}}{%
{\protect \APACyear {2014}}%
}]{%
QNLStakei2014temperature}
\APACinsertmetastar {%
QNLStakei2014temperature}%
\begin{APACrefauthors}%
Takei, Y.%
, Karasawa, F.%
\BCBL {}\ \BBA {} Yamauchi, H.%
\end{APACrefauthors}%
\unskip\
\newblock
\APACrefYearMonthDay{2014}{}{}.
\newblock
{\BBOQ}\APACrefatitle {Temperature, grain size, and chemical controls on polycrystal anelasticity over a broad frequency range extending into the seismic range} {Temperature, grain size, and chemical controls on polycrystal anelasticity over a broad frequency range extending into the seismic range}.{\BBCQ}
\newblock
\APACjournalVolNumPages{Journal of Geophysical Research: Solid Earth}{119}{7}{5414--5443}.
\PrintBackRefs{\CurrentBib}

\bibitem [\protect \citeauthoryear {%
Tan%
, Jackson%
\BCBL {}\ \BBA {} Fitz~Gerald%
}{%
Tan%
\ \protect \BOthers {.}}{%
{\protect \APACyear {2001}}%
}]{%
tan2001}
\APACinsertmetastar {%
tan2001}%
\begin{APACrefauthors}%
Tan, B.%
, Jackson, I.%
\BCBL {}\ \BBA {} Fitz~Gerald, J.%
\end{APACrefauthors}%
\unskip\
\newblock
\APACrefYearMonthDay{2001}{}{}.
\newblock
{\BBOQ}\APACrefatitle {High-temperature viscoelasticity of fine-grained polycrystalline olivine} {High-temperature viscoelasticity of fine-grained polycrystalline olivine}.{\BBCQ}
\newblock
\APACjournalVolNumPages{Physics and Chemistry of Minerals}{28}{}{641--664}.
\PrintBackRefs{\CurrentBib}

\bibitem [\protect \citeauthoryear {%
Ulloa%
, Winters%
, W{\"u}est%
\BCBL {}\ \BBA {} Bouffard%
}{%
Ulloa%
\ \protect \BOthers {.}}{%
{\protect \APACyear {2019}}%
}]{%
ulloa2019differential}
\APACinsertmetastar {%
ulloa2019differential}%
\begin{APACrefauthors}%
Ulloa, H\BPBI N.%
, Winters, K\BPBI B.%
, W{\"u}est, A.%
\BCBL {}\ \BBA {} Bouffard, D.%
\end{APACrefauthors}%
\unskip\
\newblock
\APACrefYearMonthDay{2019}{}{}.
\newblock
{\BBOQ}\APACrefatitle {Differential heating drives downslope flows that accelerate mixed-layer warming in ice-covered waters} {Differential heating drives downslope flows that accelerate mixed-layer warming in ice-covered waters}.{\BBCQ}
\newblock
\APACjournalVolNumPages{Geophysical Research Letters}{46}{23}{13872--13882}.
\PrintBackRefs{\CurrentBib}

\bibitem [\protect \citeauthoryear {%
Winters%
, Ulloa%
, W{\"u}est%
\BCBL {}\ \BBA {} Bouffard%
}{%
Winters%
\ \protect \BOthers {.}}{%
{\protect \APACyear {2019}}%
}]{%
winters2019energetics}
\APACinsertmetastar {%
winters2019energetics}%
\begin{APACrefauthors}%
Winters, K\BPBI B.%
, Ulloa, H\BPBI N.%
, W{\"u}est, A.%
\BCBL {}\ \BBA {} Bouffard, D.%
\end{APACrefauthors}%
\unskip\
\newblock
\APACrefYearMonthDay{2019}{}{}.
\newblock
{\BBOQ}\APACrefatitle {Energetics of radiatively heated ice-covered lakes} {Energetics of radiatively heated ice-covered lakes}.{\BBCQ}
\newblock
\APACjournalVolNumPages{Geophysical Research Letters}{46}{15}{8913--8925}.
\PrintBackRefs{\CurrentBib}

\bibitem [\protect \citeauthoryear {%
Zhang%
, Jiang%
, Stuecker%
, Jin%
\BCBL {}\ \BBA {} Timmermann%
}{%
Zhang%
\ \protect \BOthers {.}}{%
{\protect \APACyear {2021}}%
}]{%
zhang2021spurious}
\APACinsertmetastar {%
zhang2021spurious}%
\begin{APACrefauthors}%
Zhang, W.%
, Jiang, F.%
, Stuecker, M\BPBI F.%
, Jin, F\BHBI F.%
\BCBL {}\ \BBA {} Timmermann, A.%
\end{APACrefauthors}%
\unskip\
\newblock
\APACrefYearMonthDay{2021}{}{}.
\newblock
{\BBOQ}\APACrefatitle {Spurious north tropical Atlantic precursors to El Ni{\~n}o} {Spurious north tropical atlantic precursors to el ni{\~n}o}.{\BBCQ}
\newblock
\APACjournalVolNumPages{Nature Communications}{12}{1}{3096}.
\PrintBackRefs{\CurrentBib}

\end{thebibliography}

\end{document}